\documentstyle[epsfig,12pt]{article}
\begin{document}
\begin{center}
{\Large Relating the proton, neutron and deuteron structure

functions in the covariant Bethe-Salpeter formalism}\\[.5cm]


{V.V. Burov$^{1,2}$, A.V. Molochkov$^{2,3}$, G.I. Smirnov $^{1,4}$
and H.~Toki$^{2}$}\\[0.5cm]

$^{1}$Joint Institute for Nuclear Research, Dubna, Russia

$^{2}$Research Center for Nuclear Physics, Osaka University, Osaka, Japan

$^{3}$Far East State University, Vladivostok, Russia

$^{4}$Universit\'{e} Blaise Pascal, Clermont-Ferrand, France

\end{center}

\begin{abstract}
The neutron structure function
$F_2^{\rm n}(x)$ is evaluated within the kinematic range $10^{-3}<x<1$
from the deuteron and proton data by employing relativistic
theoretical description of $F_2^{\rm D}(x)$ and several
assumptions on the high-$x$ asymptotics of $F_2^{\rm n}(x)/F_2^{\rm p}(x)$.
It is shown that new measurements of $F_2^{\rm D}(x)$ in the range
$0.6 < x \leq 0.8$ would substantially improve understanding of the
relation between $d$ and $u$ valence quarks in the limit
$x \to 1$.
\end{abstract}

\section{Introduction}
The valence quark structure of the proton and neutron has for some
time been assumed to be understood. A number of experiments have
provided a detailed representation of the nucleon's quark distributions 
over a wide range of kinematics with some exceptions: the range of
Bjorken $x$ close to a single nucleon kinematic limit, $ x = 1$ ,
remains to be inaccessible. Actually, the situation is even more
unfortunate because the range beyond $x= 0.7$ has been poorly
explored experimentally. Recent reviews of the neutron and proton
structure, which still do not pretend to be complete, are
presented in Refs.~\cite{sciulli,mt3He}.

The valence $u$ and $d$ quark distributions are generally obtained
from measurements of the proton and neutron structure functions,
$F_2^{\rm p}(x)$ and $F_2^{\rm n}(x)$, respectively. The $u$ quark
distribution is relatively well constrained by the $F_2^{\rm p}(x)$ data
for $x \leq 0.8$ but the absence of free neutron targets has
resulted in large uncertainties in the $d$ quark distribution
beyond $x > 0.6$. Major uncertainties arise from model
considerations of the deuteron structure function $F_2^{\rm D}(x)$, from
which $F_2^{\rm n}(x)$ is evaluated. They are normally represented as a
spectrum of different ratios $R^{\rm n/p} \equiv$ $F_2^{\rm n} /F_2^{\rm p}$
as a
function of $x$. There is no theoretical constraint on the $x$
dependence of $R^{\rm n/p}$. Numerous very different  constraints on
$R^{\rm n/p}$ at the kinematic boundary $x = 1$ have been suggested in
quark models~\cite{ioffe84,feyn72} and QCD inspired
models~\cite{farrar75,bbs95}. They rely on model considerations of
the $d/u$ ratio in the limit  $x \to 1$ and neglect the
contribution from sea quarks. 
Such predictions
offer a nice testing ground for our understanding of the role
which valence quarks play in the nucleon wave function.
However, it is extremely difficult to confront the
predictions made for the limit $x \to 1$ with measured values which 
have to be extrapolated to this limit.
This explains coexistence of
numerous models of $F_2^{\rm D}(x)$ used in practice of measurements of
$F_2^{\rm n}(x)$ as well as for
motivating new experimental research into the valence quark structure.


As it is demonstrated in Ref.~\cite{bodek99} by using examples of
evaluating the $d/u$ ratio, improvement of the knowledge of
$F_2^{\rm n}(x)$ in the region $x \geq 0.75$ is very important for many
applications in hadron physics.
The procedure of extraction of 
$F_2^{\rm n}(x)$ from the $F_2^{\rm D}(x)$ data and its
theoretical justification was a subject of many publications
(see e.g. Refs.~\cite{thomas}), which defended 
a rigorous consideration of the role of nuclear effects. 
From the intense discussion in Ref.~\cite{bo-mel} one learns that
compromises and simplifications distort considerably   
$F_2^{\rm n}(x)$ and, therefore,  $d/u$ in the high $x$ range.
Theoretical understanding of nuclear effects is further discussed 
in Ref.~\cite{melni02} by presenting model independent relations which
follow from the concept of quark-hadron duality.

The present paper continues the series of
publications~\cite{bodek99} -- \cite{pace01} by suggesting an
alternative approach to extraction of the $F_2^{\rm n}(x)$ from the
data collected in deep inelastic scattering (DIS) experiments
which relies on relativistic theoretical description of
$F_2^{\rm D}(x)$~\cite{bbmst,bm98} and well defined assumptions on the
high-$x$ asymptotics for $R^{\rm n/p}$. We forward our criticism
against unjustified simplifications frequently made in
consideration of the ``nuclear'' effects in $F_2^{\rm D}$. Most common
misapprehension shows up in attempts to find an analogy (and even
extrapolation rule) between the EMC effect and a modification of
the nucleon structure inside the deuteron. There is {\em no such
theoretical concept} as the EMC effect, which is just a bare
observation that  $F_2^A /F_2^{\rm D} < 1$ in a certain range of $x$.
Therefore, there are no grounds to relate directly the difference
between  $F_2^A$ and $F_2^{\rm D}$ with that of 
 $F_2^{\rm D}$ and the free nucleon structure function $F_2^{\rm N}$,
where $F_2^{\rm N} \equiv$ $(F_2^{\rm n} + F_2^{\rm p})/2$.
The difference between $F_2^{\rm N}$ and  $F_2^{\rm D}$ 
can be conceptually
original~\cite{bms_phlett}.

Experimental information on the deuteron and proton
structure functions is available from  the
experiments of BCDMS, EMC, SLAC, E665, NMC, H1~\cite{alldata}.
The data for $F_2^{\rm D}$ and $F_2^{\rm p}$ and their ratio are available
in the range $10^{-3}<x<0.6$. There is also data at
$0.6<x<0.9$ from SLAC with relatively large errors.
The SMC collaboration proposed overall fit of world data which
gives $F^{\rm p}_2$ and $F_2^{\rm D}$ in the range
$10^{-4}<x<0.85$~\cite{smcfit}.
We make use of this approximation and extend it up to $x=1$
based on the quark counting rule for the $x\to 1$ limit of $F_2^{\rm p}$.
Further we consider extraction of $F^{\rm n}_2(x)$ by assuming that
the deuteron can be  considered as the two-nucleon bound
state.
We show that the behavior of $F^{\rm n}_2(x)$ outside the $x$ range
covered by measurements can be established by employing
well known theoretical prescriptions.
Possible  ambiguities connected with these prescriptions
are investigated.

\section{Neutron structure function}

Our consideration of the deuteron is based on the standard picture
of the proton and neutron bound together into the simplest
nucleus. It is then a favored source of information about neutron
structure function $F^{\rm n}_2(x)$. The main obstacle in the
quantitative evaluation of $F^{\rm n}_2(x)$ in this case comes
from nuclear binding 
which is neglected in many analyses on the grounds that the
deuteron binding energy is very small. Of course, one can assume that
it will be sufficient to consider just Fermi motion which
becomes particularly important at large Bjorken $x$.
On the other hand the EMC effect~\cite{EMC} and nuclear
shadowing~\cite{shadowing} show us that even small binding can
qualitatively change the observed nucleon structure.
Its effect, as it is demonstrated by the analysis in Ref.~\cite{smir99},
is clearly manifested in the entire $x$ range, including the range
of $x < 0.1$. Up to now there is no well established and
unified explanation of nuclear effects in whole range of $x$.
Numerous models used to reproduce dynamics of the
effects are not consistent. All these facts render the procedure of
$F^{\rm n}_2(x)$ evaluation very ambiguous and model dependent.

To elucidate the problem of model dependence, one needs an
approach which is less dependent on the details of dynamical nature
 of the effects and provides more general and unified picture.
We use the approach based on the covariant Bethe-Salpeter
formalism~\cite{bbmst}. It yields a good description of the ratio of
the nuclear to deuteron structure function which is shown to
be universal for all nuclei~\cite{bms_phlett,smir99}.
Within this approach the hadronic part of the nuclear deep
inelastic amplitude $W^{A}_{\mu\nu}$ is expressed in terms of the
off-mass-shell nucleon  and antinucleon amplitudes,
$W^{\rm N}_{\mu\nu}$ and $W^{\rm {\overline N}}_{\mu\nu}$,
respectively, by the following expression:
\begin{equation}
W_{\mu \nu}^A(P,q)=\sum_{i}\int dk_i
(W^{\rm N}_{\mu\nu}(k_i,q)f^{{\rm N}/A}(P,k_i)+ W^{{\rm \overline
N}}_{\mu\nu}(k_i,q)f^{{\rm\overline N}/A}(P,k_i))~,\label{hadr}
\end{equation}
where the indices $\mu\nu$ denote the Lorentz components of the
amplitude, index $i$ counts nucleons inside the nucleus,  $P$ is
the total momentum of the nucleus, $k_i$ is the relative momentum
of the struck nucleon and $q$ is the transferred momentum from the
photon. The distribution function $f^{{\rm N}/A}$ is expressed in
terms of the $n$-nucleon Bethe-Salpeter vertex functions
\begin{eqnarray}
f^{{\rm N}/A}(P,k_i)=\int dk_1\dots dk_{i-1}dk_{i+1}\dots dk_n \overline
u({\bf k}_i)S_{(n)}(P, k_i)u({\bf k}_i) \overline{\Gamma}(P,{\cal
K})S_{(n)}(P, {\cal
K})\Gamma(P,{\cal K}),\nonumber\\
f^{{\rm\overline N}/A}(P,k_i)=\int dk_1\dots dk_{i-1}dk_{i+1}\dots
dk_n \overline v({\bf k}_i)S_{(n)}(P, k_i)v({\bf k}_i)
\overline{\Gamma}(P,{\cal K})S_{(n)}(P, {\cal K})\Gamma(P,{\cal
K}). \label{distr}
\end{eqnarray}
It gives $4D$ momentum distribution of the nucleon and antinucleon
inside a nucleus. Thus, according to Eq.~(\ref{hadr}),
all nuclear effects should follow from the $4D$ Fermi motion
of the nucleon inside a nucleus~\cite{m_nucl}.
The time component of the Fermi motion is exclusive feature of the
relativistic approach. In the $3D$ limit this component results in
the change of the nucleon structure~\cite{m_nucl}.
Therefore, one can take explicit account of it by using a dynamical
model for the interlinkage of the nucleon and nuclear structure.
This problem is closely related to that of the
off-mass-shell effects. Since the nucleon amplitude
$W^{\rm N}_{\mu\nu}$ in Eq.~(\ref{hadr}) is off shell
 it cannot be connected with corresponding observable nucleon amplitude.
This point makes
Eq.~(\ref{hadr}) useless for practical applications.
 To solve this problem one can use analytic properties of the integrand
 in~(\ref{hadr}) and remove explicitly the Fermi motion along time axis
by integrating over $k_0$. 
We do it assuming small relative momenta of bound nucleons.
Preserving general form of Eq.~(\ref{hadr})
for the lightest nuclei one can transform it as follows:
\begin{eqnarray}
W_{\mu\nu}^{A}(P,q)=
\sum\limits_{a,a^\prime}^{A-1}\int\frac{d^3k_a}{(2\pi)^3}
\left[W_{\mu\nu}^{a}(k_a,q)+ \Delta^{A}_{a,a^\prime}\frac{{d}
W_{\mu\nu}^{a}(k_a,q)} {d {k_{a}}_0} \right]_{{k_a}_0=\sqrt{{\bf
k_a}^2+M_a^2}} \Phi_{a,a^\prime}^{A}({\bf k})^2, \label{hadr1}
 \end{eqnarray}
 where $a$ and $a^\prime$ denote the
struck and spectator nuclear constituent, respectively.
Depending on the mass of the considered nucleus, $a$ and $a^\prime$
assume different symbols:
\begin{eqnarray}
(a,a^\prime)= \left\{\begin{array}{rl}
 &{\rm (p,n), (n,p)}, \mbox{ if } A=2\,\, ({\rm D})\\
 &{\rm (p,pn), (n,pp), (p,D), (D,p)}, \mbox{ if } A=3\,\, ({\rm ^3He})\\
 &{\rm (n,pn), (n,pn), (n,D), (D,n)},  \mbox{ if } A=3\,\, ({\rm ^3H}) \\
 & \left. \begin{array}{rl}  &{\rm (p,pnn), (n,ppn), (p,Dn), (n,Dp),} \\
&{\rm (D,pn), (D,D), (p,^3H), (n,^3He),} \end{array}\right\}
\mbox{ if } A=4 \,\, ({\rm ^4He})~,
\end{array}\right.
\label{a_struc}\end{eqnarray}
where $\Delta^{A}_{a,a^\prime}=M_A-E_{a}-E_{a^\prime}$
is the separation energy of the corresponding nuclear fragment.
Now the nuclear effects are interpreted by the conventional $3D$ Fermi
motion of nuclear fragments and the derivative with respect to
$k_0$ of its DIS on-shell amplitudes. The distribution
function $\Phi_{a,a^\prime}^{A}({\bf k})^2$ is defined by the
projection of the Eq.~(\ref{distr}) onto $3D$ space
\begin{equation}
\Phi_{a,a^\prime}^{A}({\bf k})^2=\left\{f^{\widetilde
{\rm N}/A}(P,k_i)\right\}_{p_i^2=M_a^2}
\end{equation}
and it is closely related to the nuclear spectral function. The term
containing the derivative represents a dynamical modification of
the bound nucleon structure observed in the $3D$ projection of the
relativistic bound state as nuclear shadowing and the EMC effect.
The spectrum of bound states presented by Eq.~(\ref{a_struc}) is 
defined by analytic properties of the distribution function 
$f^{{\rm N}/A}$ and thereby ensures Pauli blocking
of the forbidden states.  
%

Starting from Eq.~(\ref{hadr1}) one can derive the
$F_2^{\rm D}$ in the form~\cite{bm98}:
\begin{eqnarray}
\hspace*{-1cm}
 F_2^{\rm D}(x_{\rm D})=\int \frac{d^3k}{(2\pi)^3}
\frac{m^2}{4E^3(M_{\rm D}-2E)^2}\left\{F_2^{\rm N}(x_{\rm N})
\left(\frac{E-k_3}{M_{\rm D}}+\frac{M_{\rm D}-2E}{M_{\rm
D}}\right) f^{\rm N/D}(M_{\rm D},k) - \right.
\nonumber\\ \label{f2d}\\
\left.
 -\frac{M_{\rm D}-2E}{M_{\rm D}} \left(x_{\rm N}\frac{dF_2^{\rm
N}(x_{\rm N})}{dx_{\rm N}}f^{\rm N/D}(M_{\rm D},k) - F_2^{\rm N}(x_{\rm
N})(E-k_3) \frac{\partial}{\partial
k_0}f^{\rm N/D}(M_{\rm D},k)\right)\right\}_{k_0=E-{\rm M_{\rm D}}/2},
\nonumber\end{eqnarray}
where $E$ is the on-mass-shell nucleon energy.
We consider this expression as an integral equation with the
unknown function $F_2^{\rm n}$. 
 Functions $F_2^{\rm p}$ and $F_2^{\rm D}$ are assumed to be known from
experiments. Since the available data does not cover the entire
kinematic range, the behaviour of these functions 
at boundaries ($x=0$, $x=1$) can not be defined from experiments.
The latter makes definition of the boundary conditions for the solution of
Eq.~(\ref{f2d})  model dependent. Basically, the boundary behavior of 
the ratio $F^{\rm n}_2/F^{\rm p}_2$ can be found from an
iteration procedure. Experience of using such procedure
 shows that the final result strongly depends on the zero
iteration~\cite{pace01}, which is hard to define 
on the basis of physical conditions.
Another way of solving this problem is to fit the right hand side of
Eq.~(\ref{f2d}) to experimental data on $F_2^{\rm D}(x)$. 
 The boundary conditions are introduced explicitly as asymptotics of the 
nucleon structure functions at $x=0$ and $x=1$ and their influence on
$F_2^{\rm D}(x)$ at medium $x$ is studied.

As an initial condition we use the following anzats for $F_2^{\rm n}(x)$:
\begin{equation}
F_2^{\rm n}(x) = R^{\rm n/p}(x)F_2^{\rm p}(x)
\label{anzats}\end{equation}
with the extended SMC fit of $F_2^{\rm p}$ and the function $R^{\rm n/p}(x)$ in
the form
\begin{equation}
R^{\rm n/p}(x)=a_1(1-x)^{\alpha_1}+a_2x^{\alpha_2}+b_1x^{\beta_1}(1-x)^{\beta_2}
(1+c_1x^{\gamma_1})~.
\label{fanzats}\end{equation}
The parameters $a_1$, $a_2$, $\alpha_1$, $b_1$ are introduced in
order to satisfy the asymptotics of the nucleon structure
functions: $a_1=1$ (corresponds to the limit $F_2^{\rm
p}(0)=F_2^{\rm n}(0)$); the parameter $a_2$ should be fixed
according to $lim_{x\to 1}F_2^{\rm n}(x)/F_2^{\rm p}(x)$. We use
three values for this limit in order to study possibility to
extract it from the experimental data. The simplest quark model
with SU(6) symmetry gives $R^{\rm n/p}_{x=1} =2/3$.
On the other hand, the kinematical limit $x=1$ corresponds
to the elastic scattering off the nucleon. Therefore,
the ratio of the nucleon structure functions becomes equivalent
to the ratio of the elastic cross sections~--- $R^{\rm n/p}_{x=1}=$
$\sigma^{\rm n}_{_{\rm elastic}}/\sigma^{\rm p}_{_{\rm elastic}}$.
It gives another value for the ratio in this limit, $R^{\rm
n/p}_{x=1} =0.47$. The minimal value of the ratio has been
provided by the model with SU(6) symmetry breaking with scalar
diquark dominance~--- $R^{\rm n/p}_{x=1} =1/4$. Thus one has a wide
range of possible values for the
structure functions ratio at $x=1$. In the next section we  show
how measurements of $F_2^{\rm D}/F_2^{\rm p}$ are sensitive to
these assumptions.
\begin{figure}[p]
\hspace*{-0.2cm}\mbox{\epsfxsize=.55\hsize\epsffile{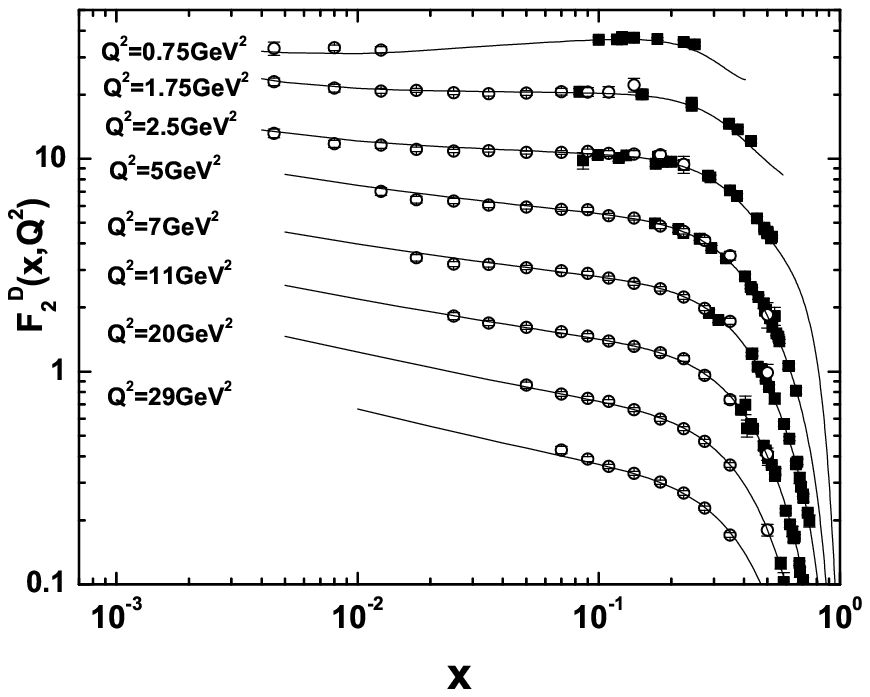}\hspace*{-1.cm}\epsfxsize=.55\hsize\epsffile{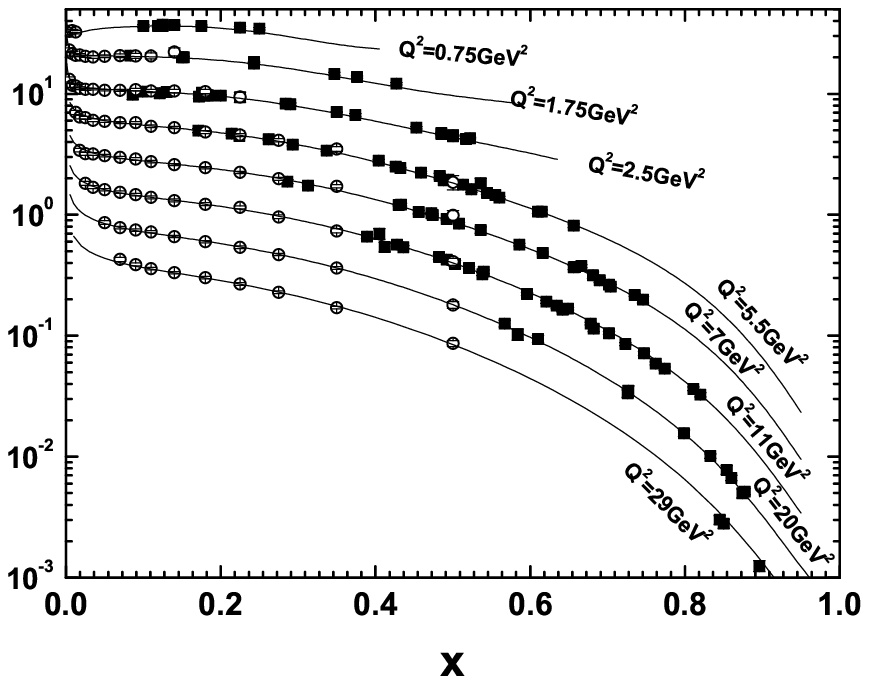}}\\[-1.cm]
\hspace*{-0.2cm}\mbox{\epsfxsize=.55\hsize\epsffile{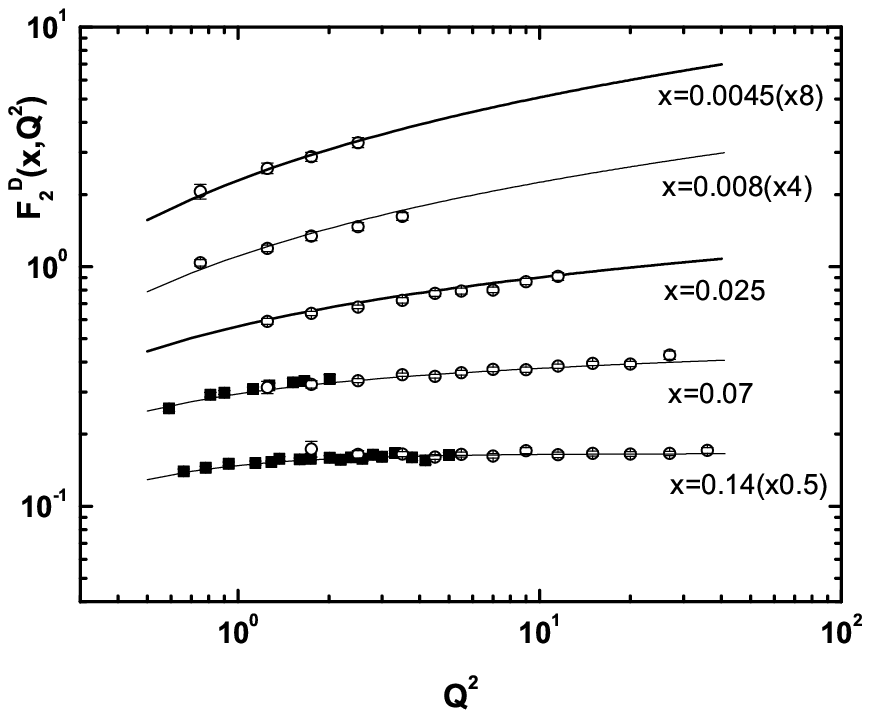}\hspace*{-1.cm}\epsfxsize=.543\hsize\epsffile{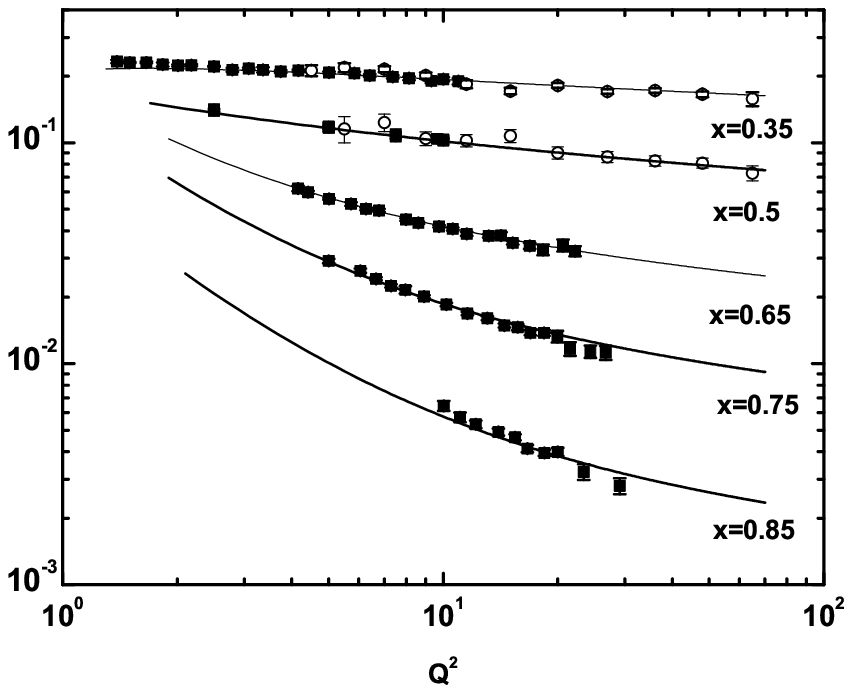}}
\caption{\label{fit} The structure function of the deuteron $F_2^{\rm D}$ 
measured in the SLAC (filled squares) and NMC (empty circles) experiments,
shown as a function of $x$ for bins of fixed $Q^2$ (left upper panel --- 
on a log scale, and right upper panel --- on a linear scale),
and as a function of $Q^2$ for bins of fixed $x$ (two lower panels).
It is approximated with Eq.~(6) in the range $10^{-3} < x < 0.6$ with the
constraints as explained in the text.
}
\end{figure}
The last constraint for the parameters can be obtained from the asymptotics 
of the proton and neutron structure functions by assuming validity
of the quark counting rule in the range $x\to 1$.
This results in similar asymptotic behavior for 
$F_2^{\rm p}(x)$  and $F_2^{\rm n}(x)$:
\begin{eqnarray}
\lim\limits_{x\to 1}F_2^{\rm p}(x)\simeq C_{\rm p}(1-x)^3, \\
\lim\limits_{x\to 1}F_2^{\rm n}(x)\simeq C_{\rm n}(1-x)^3.
\end{eqnarray}
Accordingly, the derivative of $R^{\rm n/p}$ at $x=1$ is zero because
\begin{eqnarray}
\lim\limits_{x\to 1}R^{\rm n/p}(x)=\frac{C_{\rm n}}{C_{\rm p}}=Const.
\end{eqnarray}
 This gives following constraints on the parameters of
Eq.~(\ref{fanzats}): $\alpha_1=1$, $\beta_2=1$, $b_1=(\alpha_2
a_2-1)/(1+c_1)$.
All other parameters are considered as free and used to fit
Eq.~(\ref{f2d}) to the deuteron data in the range
$10^{-3}<x<0.6$.
%
\section{Discussion of results}
The procedure described in the previous section is used to 
approximate the SLAC and NMC data on $F_2^{\rm D}(x,Q^2)$ in the
range 0.6 GeV$^2$ $\leq  Q^2 \leq $ 65 GeV$^2$ and 
$10^{-3}\leq x\leq 0.6$. This is done by fixing three
different limits of $F_2^{\rm n}/F_2^{\rm p}$ at $x=1$ and by
varying four parameters in Eq.~(\ref{fanzats}), namely
$\alpha_2$, $\beta_1$, $c_1$ and $\gamma_1$. In the considered 
kinematic range all three limits provide equally good approximation,
which means that the solution for  $F_2^{\rm n}(x)/F_2^{\rm p}(x)$
converges in the range $x < 0.6$ to the virtually unique 
function described by  Eq.~(\ref{fanzats}).
The result of the fit, which corresponds to the case
$F_2^{\rm n}(1)/F_2^{\rm p}(1)=0.47$ is displayed in Fig.~\ref{fit}.
\begin{figure}[h]
\hspace*{2cm}\mbox{\epsfxsize=.8\hsize\epsffile{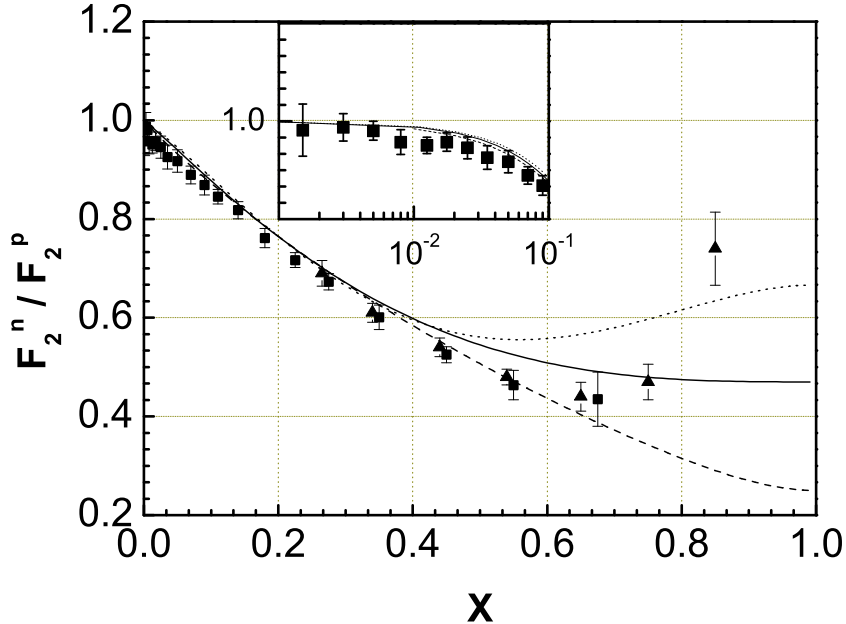}}
\vspace{-0.5cm}
\caption{\label{f2n}
The ratio of the neutron to proton structure functions evaluated
in the framework of the presented approximation by setting three
different values for the parameter $R^{\rm n/p}_{x=1}$: (1) ---
2/3 (short-dashed line), (2) --- 0.47 (solid line) 
and (3) --- 1/4 (long-dashed line).  Points with error bars
depict the results of the NMC experiment~[22] (solid squares) 
and SLAC analysis~[23] (solid triangles) obtained 
on the basis of the naive approach.
Additionally, the small $x$ range ($x < 0.1$) is displayed on a log
scale in the inset.}
\end{figure}
The three alternative solutions are shown
in Fig.~\ref{f2n} with three lines, which virtually coincide if $x < 0.4$. 
Outside this range, the obtained ratio $F_2^{\rm n}(x)/F_2^{\rm p}(x)$ is 
represented by three lines, which exactly reproduce the constraints imposed at 
$x$ = 1. 
The convergence to three 
different solutions is apparently provided by two constraints: 
available data on $F_2^{\rm D}$ and by three considered boundary conditions.
Therefore, the solution is always model independent in the range of
$x$, in which the constraint from $F_2^{\rm D}$ measurement is stronger than
 the one from boundary conditions.

The  solutions, which we find for $F_2^{\rm n}(x)$ are
compared in Fig.~\ref{f2n} with the results of NMC experiment~\cite{n-NMC}
and SLAC analysis~\cite{whitlow92}. The NMC results are obtained 
assuming a naive approach
%
%
$F_2^{\rm n}=2F_2^{\rm D}-F_2^{\rm p}$ that is equivalent to
the assumption $F_2^{\rm D}=F_2^{\rm N}$.
The points corresponding to SLAC results were transformed by us from 
$F_2^{\rm D}(x)/F_2^{\rm p}(x)$
to $F_2^{\rm n}(x)/F_2^{\rm p}(x)$ in the same naive way as 
it was done by NMC, namely by neglecting nuclear effects in the
deuteron.
In the range $x < 0.6$ all three lines are in good agreement with
the results of NMC and SLAC.
We explain the success of the naive approach in the range $x < 0.6$
by the effect of cancellation of the modifications of the free
nucleon structure function due to the nuclear (deuteron) binding
and nucleon Fermi motion in this very kinematic range. Of course,
the cancellation is not complete, which is seen from small
(2--4 \%) overestimation of the data at medium values of $x$. 
The difference between the naive approximation and the exact result 
becomes dramatic at $0.62 <x< 1$, which is a well known problem.
This is well illustrated in Fig.~\ref{f2n} by inconsistency 
between data points at $x < 0.8$ and $x > 0.8$. 
To restore the consistency, complete 
accounting of both the nuclear binding and Fermi motion has to 
be done in evaluating $R^{\rm n/p}(x)$.

\begin{figure}[h]
\hspace*{1.5cm}\mbox{\epsfxsize=.8\hsize\epsffile{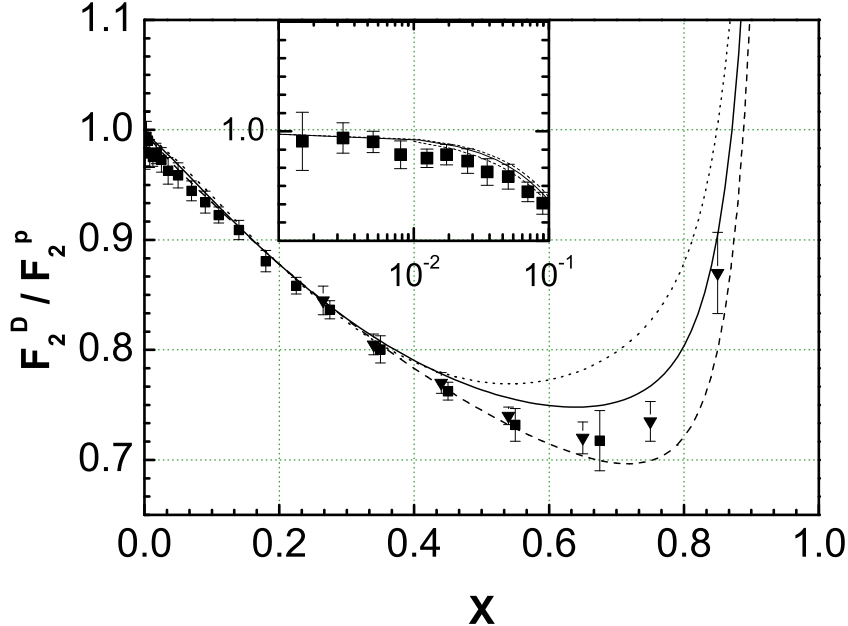}}
\vspace{-0.5cm}
\caption{\label{limits_d} The result of modification of the ratio
of the deuteron to proton structure functions evaluated for three
values of the parameter $R^{\rm n/p}_{x=1}$ as in Fig.~(2)
compared to the data of Ref.~[22] (solid squares) and results of the 
analysis performed in  Ref.~[23] (solid triangles). 
Additionally, the small $x$ range is displayed on a log
scale in the inset.}
\end{figure} 
In order to understand theoretical uncertainties in description
of the data one should
investigate how the extracted $F_2^{\rm n}(x)$ can modify theoretical
evaluation of the deuteron to proton ratio.
%
%
The three different constraints on  $R^{\rm n/p}_{x=1}$ considered
in this paper change the approximation of
 $F_2^{\rm D}(x)$ in the range of high $x$ which is better seen in
the ratio $F_2^{\rm D}/F_2^{\rm p}$ shown in Fig.~\ref{limits_d},
in which three alternative calculations are compared with
data from Ref.~\cite{n-NMC} and the results of the analysis performed
in Ref.~\cite{whitlow92}.
We conclude, that the present status both of the data and theory
does not allow to constrain the value of the
function  $R^{\rm n/p}(x)$ at $x=1$:
both 1/4 and 0.47 are in agreement with experiments, while
the value 2/3 can be regarded as less preferable.
Further understanding of the relation between  $F_2^{\rm p}(x)$
and  $F_2^{\rm n}(x)$ can not be achieved without improvement of
experimental data on $F_2^{\rm D}(x)/F_2^{\rm p}(x)$ in the range above
$x=0.6$.
The latter is also valid for points from 
Ref.~\cite{whitlow92} even if they have apparently smaller error bars. 
Contrary to the NMC data which have similar $x$ dependence 
in different $Q^2$ intervals, the data from SLAC~\cite{whitlow90} 
used for the analysis~\cite{whitlow92} scatter considerably, 
which is not consistent with the parton picture suggesting
the valence quarks dominance in the high $x$ region. 
Our understanding of the problem is that the NMC data comes from the 
{\em single} experiment, whereas SLAC results are based on 
{\em eight} experiments on deep inelastic $e-p$ and $e-d$ scattering.

We would like to note that the analysis performed in this paper
by employing Eq.~(\ref{hadr1})
is consistent with the data available for the ratio of $^4$He
and deuteron structure functions in the range
$10^{-3} < x \leq 0.8$~\cite{bbmst,bms_phlett}.
Good agreement with experiment in two different cases observed
in a wide $x$ range proves that
Eq.~(\ref{hadr1}) offers a general approach for
accounting of the nuclear effects in the whole region of $x$.

Naturally, $F_2^{\rm D}(x)$  can be further modified
in the high $x$ region if one assumes the presence
of non-nucleonic degrees of freedom in a nucleus,
which are not implied in the nucleon structure.
Moreover, if there exist  dibaryon states,
like 6$q$ states, which can not be excluded on theoretical grounds,
$F_2^{\rm D}(x)$ can significantly change at $x \simeq 1$.
Our calculations therefore provide reference lines for the
search of such effects in
forthcoming measurements proposed for the upgraded CEBAF
facility~\cite{jnaf12}.

\section{Conclusions}
We have proposed theoretically justified and fully consistent
procedure for extracting the
neutron structure function  $F_2^{\rm n}(x)$  in the
kinematic range $10^{-3} < x \leq 1$ under three different assumptions
on $F_2^{\rm n}/F_2^{\rm p}$ at  $x = 1$.
The procedure involves a 
numerical fit of the expression~(\ref{f2d}) to the deuteron data.
The performed analysis indicates that the increase in experimental
accuracy in measurements of $F_2^{\rm D}(x)/F_2^{\rm p}(x)$ in the
range $0.6 <x <0.8$ by factor of two will be sufficient for
verification of models suggested for the evaluation of the
$d/u$ ratio at $x=1$.
The developed procedure allows one to avoid appreciable
theoretical ambiguities which are present in other analyses largely
due to simplifications in the treatment of Fermi motion. 
This concerns rather wide interval of $x$ and not only high $x$ range
as it is commonly believed.
The procedure proved to be a robust one because it relies on
a good approximation of $F_2^{\rm D}(x)$ which is
not sensitive to different high $x$ limits of the neutron
structure function.
This also means that  $F_2^{\rm D}(x)$ measured
by already completed DIS experiments ($x \leq 0.9$) can be
described without introducing nonbaryonic degrees of freedom.
The interval which remains unmeasured can in principle accommodate
dibaryon states or some other exotica.
More data in the high $x$ region is required in order to obtain
model independent information on hadronic structure of the deuteron
and to find which physics is responsible for the $d/u$ ratio.

\section{Acknowledgments}
We would like to acknowledge stimulating discussions with
G.~Salm\`{e} and A.~Hosaka.  
V.B. acknowledges the support from the Russian Foundation for
Basic Research (grant RFBR 02-02-16542).
G.I.S. acknowledges the support from the Universit\'{e} Blaise 
Pascal, Clermont-Ferrand and LPC IN2P3-CNRS, France. 
V.B. and A.M. acknowledge the support and warm hospitality of the 
Research Center of Nuclear Physics, Osaka University, Osaka, Japan.

\section{Appendix}
Here we present for completeness the parametrization 
for $F_2^{\rm p}(x,Q^2)$ as suggested in Ref.~\cite{smcfit} 
and slightly modified in the present analysis:
\renewcommand{\theequation}{A.\arabic{equation}}
\setcounter{equation}{0}
\begin{eqnarray}
F_2^{\rm p}(x,Q^2)=x^{\lambda_1}(1-x)^{\lambda_2} \sum\limits_{n=1..5} C_n(1-x)^{n-1}
\left( \frac{ln(\frac{Q^2}{\Lambda})}{ln(\frac{Q_0^2}{\Lambda})}\right)^{B(x)}
\left(1+\frac{\sum\limits_{n=1..4}\kappa_nx^n}{Q^2}\right)
\end{eqnarray}
where
$$B(x)=\rho_1+\rho_2x+\frac{\rho_3}{\rho_4+x}.$$
Fit parameters as obtained in  Ref.~\cite{smcfit} are presented 
in Table~\ref{parp}. The parametrization is restricted to the kinematic
 region $3.5\cdot 10^{-5}< x <0.85$.
\begin{table}[h]
\centerline{\begin{tabular}{||c|c|c|c|c||} \hline
 $i$ & $\lambda_i$ & $\rho_i$ & $\kappa_i$ & $C_i$ \\
 \hline
 $1$ & $-0.2499713175097$ & $0.1141083888210$  & $-1.451744104784$ &
  $0.2289630236346$\\
  \hline
  $2$ & $2.396344728724$ & $-2.235597858569$  & $8.474547402342$ &
  $0.08498360257578$\\
  \hline
  $3$ & $ - $ & $0.03115195484229$  & $-34.37914208393$ &
  $3.860797992943$\\
\hline
  $4$ & $ - $ & $0.02135222381130$  & $45.88805973036$ &
  $-7.414275585348$\\
\hline
  $5$ & $ - $ & $-$  & $-$ & $3.434223579597$\\
 \hline
\end{tabular}}
\caption{Values of the parameters for $F_2^{\rm p}$ given in Eq.~(A.1).
\label{parp}}
\end{table}
 In order to extend it to the region of high $x$ and satisfy 
the $(1-x)^3$ behavior at $x\to 1$ we modify
 the parameter $\lambda_2$ as follows:
$$\lambda_2\to \tilde\lambda_2(x)=\lambda_2+(3-\lambda_2)x^{15}~.$$
This correction does not affect the values of $F_2^{\rm p}$ at $x<0.6$ 
and affords an approximation of the proton data in a much wider
kinematic region.

The neutron structure function is defined by 
Eqs.~(\ref{anzats}) and (\ref{fanzats}) in the main text. 
Taking into account the constraints on the parameters 
we arrive at the following expression for $R^{\rm n/p}$:
\begin{equation}
R^{\rm n/p}(x)=(1-x)+a_2x^{\alpha_2}+
{\alpha_2 a_2 -1 \over 1 + c_1}x^{\beta_1}(1-x) (1+c_1x^{\gamma_1})~.
\label{anzatsA}\end{equation}
The fit parameters are presented in Table~\ref{parn}.
\begin{table}[h]
\centerline{\begin{tabular}{||c|c|c|c||}
\hline & $a_2=2/3$ & $a_2=0.47$ & $a_2=1/4$\\ \hline $\alpha_2$ & $3.13971$ & $2.2262$ & $1.15416$ \\
\hline $\beta_1$ & $2.2129$ & $1.61188$ & $0.88126$ \\
\hline $c_1$ & $-1.01176$ & $-1.00692$ & $0.86217$ \\
\hline $\gamma_1$ & $0.01901$ & $0.08483$ & $5.65744$ \\
\hline
\end{tabular}}
\caption{Parameters of Eq.~(A.2) which is the constrained form
of  Eq.~(\ref{fanzats}) connecting the proton and neutron
structure functions.\label{parn}}
\end{table}


\begin{thebibliography}{999}
\vspace{0.5cm}
\bibitem{sciulli} F. Sciulli, Phil. Trans. Roy. Soc. Lond. { A359}
(2001) 241.
\vspace{-0.2cm}
\bibitem{mt3He} W. Melnitchouk and A.W. Thomas, University Adelaida
Report N$^{\circ}$ ADP-02-78/T517 and JLAB-THY-02-28, 2002.
\vspace{-0.2cm}
 \bibitem{ioffe84} B.L. Ioffe, V.A. Khoze and L.N. Lipatov,
``Hard Processes'', Volume 1, North Holland, 1984.
\vspace{-0.2cm}
\bibitem{feyn72} R.P. Feynman, ``Photon hadron interactions'',
Benjamin, New York, 1972.
\vspace{-0.2cm}
\bibitem{farrar75} G.R. Farrar and D.R. Jackson, Phys. Rev. Lett.
35 (1975) 1416.
\vspace{-0.2cm}
\bibitem{bbs95} S.J. Brodsky, M. Burkardt and I. Schmidt,
Nucl. Phys. { B} 441 (1995) 197.
\vspace{-0.2cm}
\bibitem{bodek99}
U.K. Yang, A. Bodek and Q. Fan, hep-ph/9806457 (1998);
\vspace{-0.1cm}
U.K. Yang and A. Bodek, hep-ph/9806458 (1998); 
 U.K. Yang and A. Bodek, Phys. Rev. Lett. 82 (1999) 2467.
\vspace{-0.3cm}
\bibitem{thomas} W. Melnitchouk and A.W. Thomas,  Phys. Lett. { B}
377 (1996) 11;
\vspace{-0.1cm}
 W. Melnitchouk, J. Speth and A.W. Thomas,  Phys. Lett. { B} 435
(1998) 420;
\vspace{-0.2cm}
\bibitem{bo-mel}
 W. Melnitchouk, I.R. Afnan, F. Bissey and A.W.~Thomas,
 Phys. Rev. Lett. 84 (2000) 5455;

U.K. Yang and A. Bodek, ---  {Yang and Bodek Reply}, 
Phys. Rev. Lett. 84 (2000) 5456.
\vspace{-0.2cm}
\bibitem{melni02}
W. Melnitchouk, K. Tsushima and A.W. Thomas,
Eur. Phys. J. A 14 (2002) 105.
\vspace{-0.2cm}
\bibitem{liuti95} S. Liuti and Franz Gross, Phys. Lett. { B}
356 (1995) 157.
\vspace{-0.2cm}
\bibitem{pace01} E. Pace, G. Salm\`{e}, S. Scopetta, A. Kievsky, 
Phys. Rev. C64 (2001) 055203.
\vspace{-0.2cm}
\bibitem{bbmst}  S.G.~Bondarenko, V.V.~Burov, A.V.~Molochkov,
   G.I.~Smirnov, H.~Toki,
     Prog. Part. Nucl. Phys., Vol. { 48} (2002) 449.
\vspace{-0.2cm}
\bibitem{bm98}  V.V.~Burov and A.V.~Molochkov, Nucl. Phys.
 A~637 (1998) 31.
\vspace{-0.2cm}
\bibitem{bms_phlett} V.V. Burov, A.V. Molochkov, G.I.~Smirnov, Phys. Lett.
  B~466 (1999) 1.
\vspace{-0.2cm}
\bibitem{alldata} HEPDATA, On-line Data Review,\\
http://durpdg.dur.ac.uk/hepdata/online/f2/structindex.html
\vspace{-0.2cm}
\bibitem{smcfit} SMC, B. Adeva et al., Phys. Lett. B 412 (1997) 414.
%
%
\vspace{-0.2cm}
\bibitem{EMC} EMC-NA2, J. Ashman {\em et al.},
 Phys. Lett.  B~206 (1988) {364}. 
\vspace{-0.2cm}
\bibitem{shadowing} NMC, P.~Amaudruz et al., Nucl. Phys. B~441 (1995) 3;
NMC, M.~Arneodo et al., Nucl. Phys. B~481 (1996) 23.
\vspace{-0.2cm}
\bibitem{smir99} G.I. Smirnov, Eur. Phys. J. C~10 (1999) 239.
\vspace{-0.2cm}
\bibitem{m_nucl} A.V. Molochkov,  Nucl. Phys. { A666-667} (1-4) (2000)
169.
\vspace{-0.2cm}
\bibitem{n-NMC}NMC, M. Arneodo et al., Nucl. Phys. B487 (1997) 3.
\vspace{-0.2cm}
\bibitem{whitlow92}SLAC, L.W. Whitlow et al., Phys. Lett. B282 (1992) 475.
\vspace{-0.2cm}
\bibitem{whitlow90} L.W. Whitlow, Ph.D. thesis, Stanford University,
SLAC report 357 (1990).
\vspace{-0.2cm}
\bibitem{jnaf12} Jefferson Lab Report: ``The Science Driving
the 12 GeV Uprgrade of CEBAF'', 2001.

\end{thebibliography}
\end{document}